\documentclass[11pt,letterpaper]{article}
\usepackage{fullpage, comment}
\usepackage{amsmath, amsthm}

\newtheorem{theorem}{Theorem}
\newtheorem{lemma}[theorem]{Lemma}

\let\latexcite=\cite
\def\cite{\nolinebreak\latexcite}
\let\latexref=\ref
\def\ref{\nolinebreak\latexref}

\let\epsilon=\varepsilon

\newenvironment{description*}%
  {\vspace{-2ex}
   \begin{description}%
    \setlength{\itemsep}{-1ex}%
    \setlength{\parsep}{0pt}}%
  {\end{description}}

\newenvironment{itemize*}%
  {\vspace{-2ex}
   \begin{itemize}%
    \setlength{\itemsep}{-1ex}%
    \setlength{\parsep}{0pt}}%
  {\end{itemize}
   \vspace{-1ex}}

\newcommand{\func}[1] {\texttt{#1}}

\begin{document}

\title{On Dynamic Range Reporting in One Dimension}

\author{
        Christian Worm Mortensen%
\footnote{Part of this work was done while
  the author was visiting the Max-Planck-Institut f\"ur Informatik,
  Saarbr\"ucken, as a Marie Curie doctoral fellow.} \\
\small  IT U. Copenhagen \\
        \texttt{cworm@itu.dk}
\and 
        Rasmus Pagh \\
\small  IT U. Copenhagen \\
        \texttt{pagh@itu.dk}
\and
        Mihai P\v{a}tra\c{s}cu \\
\small  MIT \\
        \texttt{mip@mit.edu}
}

\maketitle

\begin{abstract}
  We consider the problem of maintaining a dynamic set of integers and
  answering queries of the form: report a point (equivalently, all
  points) in a given interval. Range searching is a natural and
  fundamental variant of integer search, and can be solved using
  predecessor search. However, for a RAM with $w$-bit words, we show
  how to perform updates in $O(\lg w)$ time and answer queries in
  $O(\lg\lg w)$ time. The update time is identical to the van Emde
  Boas structure, but the query time is exponentially faster. Existing
  lower bounds show that achieving our query time for predecessor
  search requires doubly-exponentially slower updates. We present some
  arguments supporting the conjecture that our solution is optimal.

  Our solution is based on a new and interesting recursion idea which
  is ``more extreme'' that the van Emde Boas recursion. Whereas van
  Emde Boas uses a simple recursion (repeated halving) on each path in
  a trie, we use a nontrivial, van Emde Boas-like recursion on every
  such path. Despite this, our algorithm is quite clean when seen from
  the right angle. To achieve linear space for our data structure, we
  solve a problem which is of independent interest. We develop the
  first scheme for dynamic perfect hashing requiring sublinear
  space. This gives a dynamic Bloomier filter (an approximate storage
  scheme for sparse vectors) which uses low space. We strengthen
  previous lower bounds to show that these results are optimal.
\end{abstract}

\section{Introduction}

Our problem is to maintain a set $S$ under insertions and deletions of
values, and a range reporting query. The query $\func{findany}(a,b)$
should return an arbitrary value in $S \cap [a,b]$, or report that $S
\cap [a,b] = \emptyset$. This is a form of existential range query.
In fact, since we only consider update times above the predecessor
bound, updates can maintain a linked list of the values in $S$ in
increasing order. Given a value $x \in S \cap [a,b]$, one can traverse
this list in both directions starting from $x$ and list all values in
the interval $[a,b]$ in constant time per value.  Thus, the
$\func{findany}$ query is equivalent to one-dimensional range
reporting.

The model in which we study this problem is the word RAM. We assume
the elements of $S$ are integers that fit in a word, and let $w$ be
the number of bits in a word (thus, the ``universe size'' is $u =
2^w$). We let $n = |S|$. Our data structure will use Las Vegas
randomization (through hashing), and the bounds stated will hold with
high probability in $n$.

Range reporting is a very natural problem, and its higher-dimensional
versions have been studied for decades. In one dimension, the problem
is easily solved using predecessor search. The predecessor problem has
also been studied intensively, and the known bounds are now tight in
almost all cases \cite{beame02predecessor}. Another well-studied
problem related to ours is the lookup problem (usually solved by
hashing), which asks to find a key in a set of values. Our problem is
more general than the lookup problem, and less general than the
predecessor problem. While these two problems are often dubbed ``the
integer search problems'', we feel range reporting is an equally
natural and fundamental incarnation of this idea, and deserves similar
attention.

The first to ask whether or not range reporting is as hard as finding
predecessors were Miltersen et al in STOC'95
\cite{miltersen99asymmetric}. For the static case, they gave a data
structure with space $O(nw)$ and constant query time, which cannot be
achieved for the predecessor problem with polynomial space. An even
more surprising result from STOC'01 is due to Alstrup, Brodal and
Rauhe \cite{alstrup01range}, who gave an optimal solution for the
static case, achieving linear space and constant query time. In the
dynamic case, however, no solution better than the predecessor problem
was known. For this problem, the fastest known solution in terms of
$w$ is the classic van Emde Boas structure \cite{veb77predecessor},
which achieves $O(\lg w)$ time per operation.

For the range reporting problem, we show how to perform updates in
$O(\lg w)$ time, while supporting queries in $O(\lg\lg w)$ time. The
space usage is optimal, i.e. $O(n)$ words. The update time is
identical to the one given by the van Emde Boas structure, but the
query time is exponentially faster. In contrast, Beame and Fich
\cite[Theorem 3.7]{beame02predecessor} show that achieving any query
time that is $o(\lg w / \lg\lg w)$ for the predecessor problem
requires update time $\Omega(2^{w^{1 - \epsilon}})$, which is
doubly-exponentially slower than our update time. We also give an
interesting tradeoff between update and query times; see theorem
\ref{thm:range} below.

Our solution incorporates some basic ideas from the previous solutions
to static range reporting in one dimension
\cite{miltersen99asymmetric, alstrup01range}. However, it brings two
important technical contributions. First, we develop a new and
interesting recursion idea which is more advanced than van Emde Boas
recursion (but, nonetheless, not technically involved). We describe
this idea by first considering a simpler problem, the bit-probe
complexity of the greater-than function. Then, the solution for
dynamic range reporting is obtained by using the recursion for this
simpler problem, on \emph{every path} of a binary trie of depth
$w$. This should be contrasted to the van Emde Boas structure, which
uses a very simple recursion idea (repeated halving) on every
root-to-leaf path of the trie. The van Emde Boas recursion is
fundamental in the modern world of data structures, and has found many
unrelated applications (e.g.  exponential trees, integer sorting,
cache-oblivious layouts, interpolation search trees). It will be
interesting to see if our recursion scheme has a similar impact. 

The second important contribution of this paper is needed to achieve
linear space for our data structure. We develop a scheme for dynamic
perfect hashing, which requires sublinear space. This can be used to
store a sparse vector in small space, if we are only interested in
obtaining correct results when querying non-null positions (the
Bloomier filter problem). We also prove that our solution is
optimal. To our knowledge, this solves the last important theoretical
problem connected to Bloom filters. The stringent space requirements
that our data structure can meet are important in data-stream
algorithms and database systems. We mention one application below, but
believe others exist as well.

\subsection{Data-Stream Perfect Hashing and Bloomier Filters}

The Bloom filter is a classic data structure for testing membership in
a set. If a constant rate of false-positives is allowed, the space
\emph{in bits} can be made essentially linear in the size of the
set. Optimal bounds for this problem are obtained in
\cite{pagh05bloom}. Bloomier filters, an extension of the classical
Bloom filter with a catchy name, were defined and analyzed in the
static case by Chazelle et al \cite{chazelle04bloom}. The problem is
to represent a vector $V[0..u-1]$ with elements from $\{ 0, \dots, 2^r
- 1\}$ which is nonzero in only $n$ places (assume $n \ll u$, so the
vector is sparse). Thus, we have a sparse set as before, but with
values associated to the elements.  The information theoretic lower
bound for representing such a vector is $\Omega(n\cdot r + \lg
\binom{u}{n}) \approx \Omega(n (r + \lg u))$ bits. However, if we only
want correct answers when $V[x] \ne 0$, we can obtain a space usage of
roughly $O(nr)$ bits in the static case.

For the dynamic problem, where the values of $V$ can change
arbitrarily at any point, achieving such low space is impossible
regardless of the query and update times. Chazelle et
al.~\cite{chazelle04bloom} proved that $\Omega(n(r + \min(\lg\lg
\frac{u}{n^3}, \lg n)))$ bits are needed. No non-trivial upper bound
was known. We give matching lower and upper bounds:

\begin{theorem} \label{thm:bloomlb}
The randomized space complexity of maintaining a dynamic Bloomier
filter for $r\geq 2$ is $\Theta(n(r + \lg\lg \frac{u}{n}))$ bits in
expectation. The upper bound is achieved by a RAM data structure that
allows access to elements of the vector in worst-case constant time,
and supports updates in amortized expected $O(1)$ time.
\end{theorem}

To detect whether $V[x] = 0$ with probability of correctness at least
$1-\epsilon$, one can use a Bloom filter on top. This requires space
$\Theta(n\lg( 1/\epsilon ))$, and also works in the dynamic case
\cite{pagh05bloom}. Note that even for $\epsilon = 1$, randomization
is essential, since any deterministic solution must use $\Omega(n
\lg(u/n))$ bits of space, i.e.~it must essentially store the set of
nonzero entries in the vector.

With marginally more space, $O(n(r + \lg\lg u))$, we can make the
space and update bounds hold with high probability. To do that, we
analyze a harder problem, namely maintaining a perfect hash function
dynamically using low space. The problem is to maintain a set $S$ of
keys from $\{0, \dots, u-1\}$ under insertions and deletions, and be
able to evaluate a perfect hash function (i.e. a one-to-one function)
from $S$ to a small range. An element needs to maintain the same hash
value while it is in $S$. However, if an element is deleted and
subsequently reinserted, its hash value may change.

\begin{theorem} \label{thm:hash}
We can maintain a perfect hash function from a set $S \subset \{ 0,
\dots, u-1 \}$ with $|S| \leq n$ to a range of size $n + o(n)$, under
$n^{O(1)}$ insertions and deletions, using $O(n\lg\lg u)$ bits of
space w.h.p., plus a constant number of machine words. The function
can be evaluated in worst-case constant time, and updates take
constant time w.h.p.
\end{theorem}

This is the first dynamic perfect hash function that uses less space
than needed to store $S$ ($\lg \binom{u}{n}$ bits). Our space usage
is close to optimal, since the problem is harder than dynamic Bloomier
filtering. These operating conditions are typical of data-stream
computation, where one needs to support a stream of updates and
queries, but does not have space to hold the entire state of the data
structure. Quite remarkably, our solution can achieve this goal
without introducing errors (we use only Las Vegas randomization).

We mention an independent application of Theorem \ref{thm:hash}.
In a database we can maintain an index of a relation under insertions
of tuples, using internal memory per tuple which is logarithmic in the
length of the key for the tuple. If tuples have fixed length, they can
be placed directly in the hash table, and need only be moved if the
capacity of the hash table is exceeded.

\subsection{Tradeoffs and the scheme of things} \label{scheme}

We begin with a discussion of the greater-than problem. Consider an
infinite memory of bits, initialized to zero. Our problem has two
stages. In the update stage, the algorithm is given a number $a \in
[0..n-1]$. After seeing $a$, the algorithm is allowed to flip $O(T_u)$
bits in the memory. In the query stage, the algorithm is given a
number $b \in [0..n-1]$. Now the algorithm may inspect $O(T_q)$ bits,
and must decide whether or not $b > a$. The problem was previously
studied by Fredman \cite{fredman82sums}, who showed that $\max(T_u,
T_q) = \Omega(\lg n / \lg\lg n)$. It is quite tempting to believe that
one cannot improve past the trivial upper bound $T_u = T_q = O(\lg
n)$, since, in some sense, this is the complexity of ``writing down''
$a$. However, as we show in this paper, Fredman's bound is optimal, in
the sense that it is a point on our tradeoff curve. We give upper and
lower bounds that completely characterize the possible asymptotic
tradeoffs:

\begin{theorem} \label{thm:bitgt}
  The bit-probe complexity of the greater-than function satisfies the
  tight tradeoffs:
  
  \vspace{-4ex}
  \begin{eqnarray*}
   T_q \geq \lg\lg n,\ T_u \leq \lg n &:& T_u = \Theta(\lg_{T_q} n) \\
   T_q \leq \lg\lg n,\ T_u \geq \lg n &:& 2^{T_q} = \Theta(\lg_{T_u} n) \\
  \end{eqnarray*}
\end{theorem}
\vspace{-3ex}

As mentioned already, we use the same recursion idea as in the
previous algorithm for dynamic range reporting, except that we apply
this recursion to every root-to-leaf path of a binary trie of depth
$w$. Quite remarkably, these structures can be made to overlap
in-as-much as the paths overlap, so only one update suffices for all
paths going through a node. Due to this close relation, we view the
lower bounds for the greater-than function as giving an indication
that our range reporting data structure is likewise optimal. In any
case, the lower bounds show that markedly different ideas would be
necessary to improve our solution for range reporting.

Let $T_{pred}$ be the time needed by one update and one query in the
dynamic predecessor problem. The following theorem summarizes our
results for dynamic range reporting:

\begin{theorem} \label{thm:range}
  There is a data structure for the dynamic range reporting problem,
  which uses $O(n)$ space and supports updates in time $O(T_u)$, and
  queries in time $O(T_q)$, $(\forall) T_u, T_q$ satisfying:

  \vspace{-3ex}
  \begin{eqnarray*}
    T_q \geq \lg\lg w,\ \frac{\lg w}{\lg\lg w} \leq T_u \leq \lg w
      &:& T_u = O(\lg_{T_q} w) + T_{pred} \\
    T_q \leq \lg\lg w,\phantom{\ \ \frac{\lg w}{\lg\lg w} \leq} 
      T_u \geq \lg w &:& 2^{T_q} = O(\lg_{T_u} w) \\
  \end{eqnarray*}
\end{theorem}
\vspace{-3ex}

Notice that the most appealing point of the tradeoff is the cross-over
of the two curves: $T_u = O(\lg w)$ and $T_q = O(\lg\lg w)$ (and
indeed, this has been the focus of our discussion). Another
interesting point is at constant query time. In this case, our data
structure needs $O(w^{\epsilon})$ update time. Thus, our data
structure can be used as an optimal static data structure, which is
constructed in time $O(n w^{\epsilon})$, improving on the construction
time of $O(n \sqrt{w})$ given by Alstrup et al \cite{alstrup01range}.

The first branch of our tradeoff is not interesting with $T_{pred} =
\Theta(\lg w)$. However, it is generally believed that one can achieve
$T_{pred} = \Theta( \lg w / \lg\lg w)$, matching the optimal bound for
the static case. If this is true, the $T_{pred}$ term can be ignored.
In this case, we can remark a very interesting relation between our
problem and the predecessor problem. When $T_u = T_q$, the bounds we
achieve are identical to the ones for the predecessor problem, i.e.
$T_u = T_q = O(\lg w / \lg\lg w)$. However, if we are interested in
the possible tradeoffs, the gap between range reporting and the
predecessor problem quickly becomes huge. The same situation appears
to be true for deterministic dictionaries with linear space, though
the known tradeoffs are not as general as ours. We set forth the bold
conjecture (the proof of which requires many missing pieces) that all
three search problems are united by an optimal time of $\Theta(\lg w /
\lg\lg w)$ in this point of their tradeoff curves.

We can achieve bounds in terms of $n$, rather than $w$, by the classic
trick of using our structure for small $w$ and a fusion tree structure
\cite{fredman93fusion} for large $w$. In particular, we can achieve
$T_q = O(\lg\lg n)$ and $T_u = O\left( \frac{\lg n}{\lg\lg n}
\right)$. Compared with the optimal bound for the predecessor problem
of $\Theta\left( \sqrt{\frac{\lg n}{\lg\lg n}} \right)$, our data
structure improves the query time exponentially by sacrificing the
update time quadratically.

\section{Data-Stream Perfect Hashing}

We denote by $S$ be the set of values that we need to hash at present
time. Our data structure has the following parts:

\begin{itemize}
\item A hash function $\rho: \{0,\dots,u-1 \} \rightarrow
  \{0,1\}^{v}$, where $v = O(\lg n)$, from a family of universal hash
  functions with small representations (for example, the one
  from \cite{dietzfel96universal}).

\item A hash function $\phi: \{0,1\}^{v} \rightarrow \{1,\dots,r\}$,
  where $r=\lceil n/\lg^2 n \rceil$, taken from Siegel's class of
  highly independent hash functions \cite{siegel04hash}. 
  
\item An array of hash functions $h_1,\dots,h_r: \{0,1\}^v \rightarrow
  \{0,1\}^s$, where $s=\lceil (6+2c)\lg\lg u \rceil$, chosen
  independently from a family of universal hash functions; $c$ is a
  constant specified below.
  
\item A high performance dictionary \cite{dietzfel90highperf} for a
  subset $S'$ of the keys in $S$. The dictionary should have a
  capacity of $O(\lceil n/\lg u \rceil)$ keys (but might expand
  further). Along with the dictionary we store a linked list of length
  $O(\lceil n/\lg u \rceil)$, specifying certain vacant positions in
  the hash table.
  
\item An array of dictionaries $D_1,\dots,D_r$, where $D_i$ is a
  dictionary that holds $h_i(\rho(k))$ for each key $k\in S \setminus
  S'$ with $\phi(\rho(k))=i$. A unique value in $\{0,\dots,j-1\}$,
  where $j=(1+o(1))\lg^2 n$, is associated with each key in $D_i$. A
  bit vector of $j$ bits and an additional string of $\lg n$ bits is
  used to keep track of which associated values are in use. We will
  return to the exact choice of $j$ and the implementation of the
  dictionaries.
\end{itemize}

The main idea is that all dictionaries in the construction assign to
each of their keys a unique value within a subinterval of $[1 .. m]$.
Each of the dictionaries $D_1, \dots, D_r$ is responsible for an
interval of size $j$, and the high performance dictionary is
responsible for an interval of size $O(n/\lg u) = o(n)$.

The hash function $\rho$ is used to reduce the key length to $v$. The
constant in $v = O(\lg n)$ can be chosen such that with high
probability, over a polynomially bounded sequence of updates, $\rho$
will never map two elements of $S$ to the same value (the conflicts,
if they occur, end up in $S'$ and are handled by the high performance
dictionary).

When inserting a new value $k$, the new key is included in $S'$ if
either:

\begin{itemize}
\item There are $j$ keys in $D_i$, where $i=\phi(\rho(k))$, or
  
\item There exists a key $k'\in S$ where
  $\phi(\rho(k))=\phi(\rho(k'))=i$ and $h_i(\rho(k))=h_i(\rho(k'))$.
\end{itemize}

Otherwise $k$ is associated with the key $h_i(\rho(k))$ in $D_i$.
Deletion of a key $k$ is done in $S'$ if $k\in S'$, and otherwise the
associated key in the appropriate $D_i$ is deleted.

To evaluate the perfect hash function on a key $k$ we first see
whether $k$ is in the high performance dictionary. If so, we return
the value associated with $k$. Otherwise we compute $i=\phi(\rho(k))$
and look up the value $\Delta$ associated with the key $h_i(\rho(k))$
in $D_i$. Then we return $(i-1)j+\Delta$, i.e., position $\Delta$
within the $i$-th interval.

Since $D_1,\dots,D_r$ store keys and associated values of $O(\lg\lg
u)$ bits, they can be efficiently implemented as constant depth search
trees of degree $w^{\Omega(1)}$, where each internal node resides in a
single machine word. This yields constant time for dictionary
insertions and lookups, with an optimal space usage of $O(\lg^2
n\lg\lg u)$ bits for each dictionary.  We do not go into details of
the implementation as they are standard; refer to \cite{hagerup98ram}
for explanation of the required word-level parallelism techniques.

What remains to describe is how the dictionaries keep track of vacant
positions in the hash table in constant time per insertion and
deletion. The high performance dictionary simply keeps a linked list
of all vacant positions in its interval. Each of $D_1,\dots,D_r$
maintain a bit vector indicating vacant positions, and additional
$O(\lg n)$ summary bits, each taking the or of an interval of size
$O(\lg n)$. This can be maintained in constant time per operation,
employing standard techniques.

Only $o(n)$ preprocessing is necessary for the data structure
(essentially to build tables needed for the word-level parallelism).
The major part of the data structure is initialized lazily.

\subsection{Analysis}

Since evaluation of all involved hash functions and lookup in the
dictionaries takes constant time, evaluation of the perfect hash
function is done in constant time. As we will see below, the high
performance dictionary is empty with high probability unless $n/\lg u
> \sqrt{n}$. This means that it always uses constant time per
update with high probability in $n$. All other operations done for
update are easily seen to require constant time w.h.p.

We now consider the space usage of our scheme. The function $\rho$ can
be represented in $O(w)$ bits. Siegel's highly independent hash
function uses $o(n)$ bits of space. The hash functions $h_1,\dots,h_r$
use $O(\lg n + \lg\lg u)$ bits each, and $o(n\lg\lg u)$ bits in
total.  The main space bottleneck is the space for $D_1,\dots,D_r$,
which sums to $O(n\lg\lg u)$.

Finally, we show that the space used by the high performance
dictionary is $O(n)$ bits w.h.p. This is done by showing that each of
the following hold with high probability throughout a polynomial
sequence of operations:

\begin{itemize*}
\item[1.] The function $\rho$ is one-to-one on $S$.
  
\item[2.] There is no $i$ such that $S_i = \{ k \in S \mid
  \phi(\rho(k))=i \}$ has more than $j$ elements.

\item[3.] The set $S'$ has $O(\lceil n/\lg u \rceil)$ elements.
\end{itemize*}

That 1.~holds with high probability is well known. To show 2.~we use
the fact that, with high probability, Siegel's hash function is
independent on every set of $n^{\Omega(1)}$ keys. We may thus employ
Chernoff bounds for random variables with limited independence to
bound the probability that any $i$ has $|S_i| > j$, conditioned on the
fact that 1.~holds. Specifically, we can use \cite[Theorem
5.I.b]{schmidt95chernoff} to argue that for any $l$, the probability
that $|S_{i}| > j$ for $j = \lceil \lg^2 n + \lg^{5/3} n \rceil$ is
$n^{-\omega(1)}$, which is negligible. On the assumption that 1.~and
2.~hold, we finally consider~3. We note that every key $k'\in S'$ is
involved in an $h_i$-collision in $S_i$ for $i=\phi(\rho(k'))$,
i.e.~there exists $k''\in S_i \setminus \{k'\}$ where
$h_i(k')=h_i(k'')$. By universality, for any $i$ the expected number
of $h_i$-collisions in $S_i$ is $O(\lg^4 n / (\lg u)^{6+2c}) = O((\lg
u)^{-(2+2c)})$.  Thus the probability of one or more collisions is
$O((\lg u)^{-(2+2c)})$.  For $\lg u \geq \sqrt{n}$ this means that
there are no keys in $S'$ with high probability. Specifically, $c$ may
be chosen as the sum of the constants in the exponents of the length
of the operation sequence and the desired high probability bound. For
the case $\lg u < \sqrt{n}$ we note that the expected number of
elements in $S'$ is certainly $O(n/\lg u)$. To see than this also
holds with high probability, note that the event that one or more keys
from $S_i$ end up in $S'$ is independent among the $i$'s. Thus we can
use Chernoff bounds to get that the deviation from the expectation is
small with high probability.

\section{Lower Bound for Bloomier Filters}

For the purpose of the lower bound, we consider the following two-set
distinction problem, following \cite{chazelle04bloom}. The problem has
the following stages:

\begin{enumerate}
\item[0.] a random string $R$ is drawn, which will be available to the
  data structure throughout its operation. This is equivalent to
  drawing a deterministic algorithm from a given distribution, and is
  more general than assuming each stage has its own random coins (we
  are giving the data structure free storage for its random bits).

\item the data structure is given $A \subset [u], |A| \le n$. It must
  produce a representation $f_R(A)$, which for any $A$ has size at
  most $S$ bits, in expectation over all choices of $R$. Here $S$ is a
  function of $n$ and $u$, which is the target of our lower bound.

\item the data structure is given $B \subset [u]$, such that $|B| \le
  n, A \cap B = \emptyset$. Based on the old state $f_R(A)$, it must
  produce $g_R(B, f_R(A))$ with expected size at most $S$ bits.

\item the data structure is given $x \in [u]$ and its previously
  generated state, i.e.~$f_R(A)$ and $g_R(B, f_R(A))$. Now it must
  answer as follows with no error allowed: if $x \in A$, it must
  answer ``A''; if $x \in B$, it must answer ``B''; if $x \notin A
  \cup B$, it can answer either ``A'' or ``B''. Let $h_R(x,f,g)$
  be the answer computed by the data structure, when the previous
  state is $f$ and $g$.

\end{enumerate}

It is easy to see that a solution for dynamic Bloomier filters
supporting ternary associated data, using expected space $o(n\lg\lg
\frac{u}{n})$, yields a solution to the two-set distinction problem
with $S = o(n\lg\lg \frac{u}{n})$. We will prove such a solution does
not exist.

Since a solution to the distinction problem is not allowed to make an
error we can assume w.l.o.g.~that step 3 is implemented as follows. If
there exist appropriate $A, B \subset [u]$, with $x \in A$ such that
$f_R(A) = f_0$ and $g_R(B, f_0) = g_0$, then $h_R(x, f_0, g_0)$ must
be ``A''. Similarly, if there exists a plausible scenario with $x \in
B$, the answer must be ``B''. Otherwise, the answer can be arbitrary.

Assume that the inputs $A \times B$ are drawn from a given
distribution. We argue that if the expected sizes of $f$ and $g$ are
allowed to be at most $2S$, the data structure need not be
randomized. This uses a bicriteria minimax principle. We have
$E_{R,A,B}\left[ \frac{|f|}{S} + \frac{|g|}{S} \right] \leq 2$, where
$|f|, |g|$ denote the length of the representations. Then, there
exists a random string $R_0$ such that $E_{A,B} \left[ \frac{|f|}{S} +
\frac{|g|}{S} \right] \leq 2$. Since $|f|, |g| \geq 0$, this implies
$E_{A,B}[|f|] \leq 2S, E_{A,B}[|g|] \leq S$. The data structure can
simply use the deterministic sequence $R_0$ as its random bits; we
drop the subscript from $f_R, g_R$ when talking about this
deterministic data structure.

\subsection{Lower Bound for Two-Set Distinction}

Assume $u = \omega(n)$, since a lower bound of $\Omega(n)$ is trivial
for universe $u \ge 2n$. Break the universe into $n$ equal parts $U_1,
\dots, U_n$; w.l.o.g.~assume $n$ divides $u$, so $|U_i| =
\frac{n}{u}$. The hard input distribution chooses $A$ uniformly at
random from $U_1 \times \dots \times U_n$. We write $A = \{ a_1,
\dots, a_n \}$, where $a_i$ is a random variable drawn from
$U_i$. Then, $B'$ is chosen uniformly at random from the same product
space; again $B' = \{b_1, \dots, b_n\}, b_i \gets U_i$. We let $B = B'
\setminus A$. Note that $E[|B|] = n \cdot \Pr[A_1 \ne B_1] = (1 -
\frac{n}{u}) \cdot n = (1 - o(1)) \cdot n$.

Let $A_i^p$ be the plausible values of $A_i$ after we see $f(A)$; that
is, $A_i^p$ contains all $a \in U_i$ for which there exists a valid
$A'$ with $a \in A'$ and $f(A') = f(A)$. Intuitively speaking, if
$f(A)$ has expected size $o(n \lg\lg \frac{u}{n})$, it contains on
average $o(\lg\lg \frac{u}{n})$ bits of information about each
$a_i$. This is much smaller than the range of $a_i$, which is
$\frac{u}{n}$, so we would expect that the average $|A_i^p|$ is quite
large, around $\frac{u}{n} / (\lg \frac{u}{n})^{o(1)}$. This intuition
is formalized in the following lemma:

\begin{lemma}
With probability at least a half over a uniform choice of $A$ and $i$,
we have $|A_i^p| \geq \frac{u/n}{2^{O(S/n)}}$.
\end{lemma}

\begin{proof}
The Kolmogorov complexity of $A$ is $n\lg \frac{u}{n} - O(1)$; no
encoding for $A$ can have an expected size less than this quantity.
We propose an encoding for $A$ consisting of two parts: first, we
include $f(A)$; second, for each $i$ we include the index of $a_i$ in
$|A_i^p|$, using $\lceil \lg|A_i^p| \rceil$ bits. This is easily
decodable. We first generate all possible $A'$ with $f(A') = f(A)$,
and thus obtain the sets $A_i^p$. Then, we extract from each plausible
set the element with the given index. The expected size of the
encoding is $2S + \sum_i E_{A}[\lg |A_i^p|] + O(n)$, which must be
$\ge n\lg \frac{u}{n} - O(1)$. This implies $\lg \frac{u}{n} -
E_{i,A}[\lg |A_i^p|] \le \frac{2S}{n} + O(1)$. By Markov's inequality,
with probability at least a half over $i$ and $A$, $\lg \frac{u}{n} -
\lg |A_i^p| \le \frac{4S}{n} + O(1)$, so $\lg |A_i^p| \ge \lg
\frac{u}{n} - O(\frac{S}{n})$.
\end{proof}

We now make a crucial observation which justifies our interest in
$A_i^p$. Assume that $b_i \in A_i^p$. In this case, the data structure
must be able to determine $b_i$ from $f(A)$ and $g(B,f(A))$. Indeed,
suppose we compute $h(x,f,g)$ for all $x \in |A_i^p|$. If that data
strucuture does not answer ``B'' when $x = b_i$, it is obviously
incorrent. On the other hand, if it answers ``B'' for both $x = b_i$
and some other $x' \in A_i^p$, it also makes an error. Since $x'$ is
plausible, there exist $A'$ with $x' \in A'$ such that $f(A') =
f(A)$. Then, we can run the data structure with $A'$ as the first set
and $B$ as the second set. Since $f(A') = f(A)$, the data structure
will behave exactly the same, and will incorrectly answer ``B'' for
$x'$.

To draw our conclusion, we consider another encoding argument, this
time in connection to the set $B'$. The Kolmogorov complexity of $B'$
is $n \lg \frac{u}{n} - O(1)$. Consider a randomized encoding,
depending on a set $A$ drawn at random. First, we encode an $n$-bit
vector specifying which indices $i$ have $a_i = b_i$. It remains to
encode $B' \setminus A = B$. We encode another $n$-bit vector,
specifying for which positions $i$ we have $b_i \in A_i^p$. For each
$b_i \notin A_i^p$, we simply encode $B_i$ using $\lceil \lg
\frac{u}{n} \rceil$ bits. Finally, we include in the encoding $g(B,
f(A))$. As explained already, this is enough to recover all $b_i$
which are in $A_i^p$. Note that we do not need to encode $f(A)$, since
this depends only on our random coins, and the decoding algorithm can
reconstruct it.

The expected size of this encoding will be $O(n + S) + n\cdot
\Pr_{A,B',i} [b_i \notin A_i^p] \cdot \lg \frac{u}{n}$. We know that
with probability a half over $A$ and $i$, we have $|A_i^p| \geq
\frac{u/n}{2^{O(S/n)}}$. Thus, $\Pr_{A,B',i} [b_i \in A_i^p] \geq
\frac{1}{2} \cdot 2^{-O(S/n)}$. Thus, the expected size of the
encoding is at most $O(n + S) + (1 - 2^{-O(S/n)}) \cdot n \lg
\frac{u}{n}$. Note that by the minimax principle, randomness in the
encoding is unessential and we can always fix $A$ guaranteeing the
same encoding size, in expectation over $B$. We now get the bound:

\begin{eqnarray*}
& & O(n + S) + (1 - 2^{-O(S/n)}) \cdot n \lg \frac{u}{n} \geq n \lg
\frac{u}{n} - O(1) \\ 
& \Rightarrow & O\left( \frac{S}{n} \right) \geq 2^{-O(S/n)} \lg
\frac{u}{n} - O(1) \Rightarrow 2^{O(S/n)} O(S / n) \geq \lg
\frac{u}{n} \Rightarrow \frac{S}{n} = \Omega \left( \lg\lg \frac{u}{n}
\right)
\end{eqnarray*}

\section{A Space-Optimal Bloomier Filter}

It was shown in \cite{carter78bloom} that the approximate membership
problem (i.e., the problem solved by Bloom filters) can be solved
optimally using a reduction to the exact membership problem. The
reduction uses universal hashing.  In this section we extend this idea
to achieve optimal dynamic Bloomier filters.

Recall that Bloomier filters encode sparse vectors with entries from
$\{0,\dots,2^r - 1\}$.  Let $S\subseteq [u]$ be the set of at most $n$
indexes of nonzero entries in the vector $V$.  The data structure must
encode a vector $V'$ that agrees with $V$ on indexes in $S$, and such
that for any $x\not\in S$, $\Pr[V'[x]\neq 0]\leq \epsilon$, where
$\epsilon > 0$ is the error probability of the Bloomier
filter. Updates to $V$ are done using the following operations:
\begin{itemize}
\item {\sc Insert($x$, $a$)}. Set $V[x]:=a$, where $a\neq 0$.
\item {\sc Delete($x$)}. Set $V[x]:=0$.
\end{itemize}

The data structure assumes that only valid updates are performed,
i.e. that inserts are done only in situations where $V[x]=0$ and
deletions are done only when $V[x]\neq 0$.

\begin{theorem}\label{thm:filter}
Let positive integers $n$ and $r$, and $\epsilon > 0$ be given. On a
RAM with word length $w$ we can maintain a Bloomier filter $V'$ for a
vector $V$ of length $u=2^{O(w)}$ with at most $n$ nonzero entries
from $\{0,\dots,2^r - 1\}$, such that:

\begin{itemize}
\item {\sc Insert} and {\sc Delete} can be done in amortized
  expected constant time. The data structure assumes all updates are
  valid.

\item Computing $V'[x]$ on input $x$ takes worst case constant
  time. If $V[x]\neq 0$ the answer is always 'V[x]'. If $V[x]=0$ the
  answer is '0' with probability at least $1-\epsilon$.

\item The expected space usage is $O(n(\lg\lg(u/n) + \lg(1/\epsilon) +
  r))$ bits.
\end{itemize}
\end{theorem}

\subsection{The Data Structure}

Assume without loss of generality that $u\geq 2n$ and that
$\epsilon\geq u/n$.  Let $v=\max(n \log(u/n), n/\epsilon)$, and choose
$h: \{0,\dots,u-1\} \rightarrow \{0,\dots,v-1\}$ as a random function
from a universal class of hash functions. The data structure maintains
information about a minimal set $S'$ such that $h$ is 1-1 on $S
\setminus S'$. Specifically, it consists of two parts:

\begin{enumerate}
\item A dictionary for the set $S'$, with corresponding values of $V$
  as associated information.

\item A dictionary for the set $h(S\backslash S')$, where the element
  $h(x)$, $x\in S\backslash S'$, has $V[x]$ as associated information.
\end{enumerate}

Both dictionaries should succinct, i.e., use space close to the
information theoretic lower bound.  Raman and Rao
\cite{raman03succinct} have described such a dictionary using space
that is $1+o(1)$ times the minimum possible, while supporting lookups
in $O(1)$ time and updates in expected amortized $O(1)$ time.

To compute $V'[x]$ we first check whether $x\in S'$, in which case
$V'[x]$ is stored in the first dictionary. If this is not the case, we
check whether $h(x)\in h(S\backslash S')$.  If this is the case we
return the information associated with $h(x)$ in the second
dictionary.  Otherwise, we return '0'.

{\sc Insert($x$, $a$)}. First determine whether $h(x)\in h(S\backslash
S')$, in which case we add $x$ to the set $S'$, inserting $x$ in the
first dictionary.  Otherwise we add $h(x)$ to the second
dictionary. In both cases, we associate $a$ with the inserted element.

{\sc Delete($x$)} proceeds by deleting $x$ from the first dictionary
if $x\in S'$, and otherwise deleting $h(x)$ from the second
dictionary.

\subsection{Analysis}

It is easy to see that the data structure always return correct
function values on elements in $S$, given that all updates are
valid. When computing $V'[x]$ for $x\not\in S$ we get a nonzero result
if and only if there exists $x'\in S$ such that $h(x)=h(x')$. Since
$h$ was chosen from a universal family, this happens with probability
at most $n/v \leq \epsilon$.

It remains to analyze the space usage. Using once again that $h$ was
chosen from a universal family, the expected size of $S'$ is
$O(n/\log(u/n))$. This implies that the expected number of bits
necessary to store the set $S'$ is $\log\binom{u}{O(n/\log(u/n))} =
O(n)$, using convexity of the function $x\mapsto \binom{u}{x}$ in the
interval $0\dots u/2$. In particular, the first dictionary achieves an
expected space usage of $O(n)$ bits.  The information theoretical
minimum space for the set $h(S\backslash S')$ is bounded by
$\log\binom{r}{n} = O(n \log(r/n)) = O(n \log\log(u/n) +
n\log(1/\epsilon))$ bits, matching the lower bound.  We disregarded is
the space for the universal hash function, which is $O(\log u)$ bits.
However, this can be reduced to $O(\log n + \log\log u)$ bits, which
is vanishing, by using slightly weaker universal functions and
doubling the size $r$ of the range. Specifically, $2$-universal
functions suffice; see \cite{pagh00dispers} for a construction. Using
such a family requires preprocessing time $(\log u)^{O(1)}$, expected.

\section{Upper Bounds for the Greater-Than Problem}

We start with a simple upper bound of $T_u = O(\lg n), T_q = O(\lg\lg
n)$. Our upper bound uses a trie structure. We consider a balanced
tree with branching factor 2, and with $n$ leaves. Every possible
value of the update parameter $a$ is represented by a root-to-leaf
path. In the update stage, we mark this root-to-leaf path, taking time
$O(\lg n)$. In the query stage, we want to find the point where $b$'s
path in the trie would diverge from $a$'s path. This uses binary
search on the $\lg n$ levels, as follows. To test if the paths diverge
on a level, we examine the node on that level on $b$'s path.  If the
node is marked, the paths diverge below; otherwise they diverge
above. Once we have found the divergence point, we know that the
larger of $a$ and $b$ is the one following the right child of the
lowest common ancestor.

For the full tradeoff, we consider a balanced tree with branching
factor $B$. In the update stage, we need to mark a root-to-leaf path,
taking time $\lg_B n$. In the query stage, we first find the point
where $b$'s path in the trie would diverge from $a$'s path. This uses
binary search on the $\lg_B n$ levels, so it takes time $O(\lg\lg_B
n)$. Now we know the level where the paths of $a$ and $b$ diverge. The
nodes on that level from the paths of $a$ and $b$ must be siblings in
the tree. To test whether $b > a$, we must find the relative order of
the two sibling nodes. There are two strategies for this, giving the
two branches of the tradeoff curve. To achieve small update time, we
can do all work at query time. We simply test all siblings to the left
of $b$'s path on the level of divergence. If we find a marked one,
then $a$'s path goes to the left of $b$'s path, so $a < b$; otherwise
$a > b$. This stragegy gives $T_u = O(\lg_B n)$ and $T_q = O(\lg(\lg_B
n) + B)$, for any $B \geq 2$. For $T_q > \Omega(\lg\lg n)$, we have
$T_q = \Theta(B)$, so we have achieved the tradeoff $T_u = O(\lg_{T_q}
n)$.

The second strategy is to do all work at update time. For every node
on $a$'s path, we mark all left siblings of the node as such. Then to
determine if $b$'s path is to the left or to the right of $a$'s path,
we can simply query the node on $b$'s path just below the divergence
point, and see if it is marked as a left sibling. This strategy gives
$T_u = O(B \lg_B n)$ and $T_q = O(\lg(\lg_B n))$. For small enough $B$
(say $B = O(\lg n)$), this strategy gives $T_q = O(\lg\lg n)$
regardless of $B$ and $T_u$. For $B = \Omega(\lg n)$, we have $\lg B =
\Theta(\lg T_u)$. Therefore, we can express our tradeoff as: $2^{T_q}
= O(\lg_{T_u} n)$.

\section{Dynamic Range Reporting}

We begin with the case $T_u = O(\lg w), T_q = O(\lg\lg w)$. Let $S$ be
the current set of values stored by the data structure.  Without loss
of generality, assume $w$ is a power of two.  For an arbitrary $t \in
[0, \lg w]$, we define the trie of order $t$, denoted $T_t$, to be the
trie of depth $w / 2^t$ and alphabet of $2^t$ \emph{bits}, which
represents all numbers in $S$. We call $T_0$ the \emph{primary trie}
(this is the classic binary trie with elements from $S$). Observe that
we can assign distinct names of $O(w)$ bits to all nodes in all
tries. We call \emph{active paths} the paths in the tries which
correspond to elements of $S$. A node $v$ from $T_t$ corresponds to a
subtree of depth $2^t$ in the primary trie; we denote the root of this
subtree by $r_0(v)$. A node from $T_t$ corresponds to a 2-level
subtree in $T_{t-1}$; we call such a subtree a \emph{natural
subtree}. Alternatively, a 2-level subtree of any trie is natural iff
its root is at an even depth.

A root-to-leaf path in the primary trie is seen as the leaves of the
tree used for the greater-than problem. The paths from the primary
trie are broken into chunks of length $2^t$ in the trie of order
$t$. So $T_t$ is similar to the $t$-th level (counted bottom-up) of
the greater-than tree. Indeed, every node on the $t$-th level of that
tree held information about a subtree with $2^t$ leaves; here one edge
in $T_t$ summarizes a segment of length $2^t$ bits.  Also, a natural
subtree corresponds to two siblings in the greater-than structure. On
the next level, the two siblings are contracted into a node; in the
trie of higher order, a natural subtree is also contracted into a
node. It will be very useful for the reader to hold these parallels in
mind, and realize that the data structure from this section is
implementing the old recursion idea \emph{on every path}.

The root-to-leaf paths corresponding to the values in $S$ determine at
most $n-1$ branching nodes in any trie. By convention, we always
consider roots to be branching nodes. For every branching node from
$T_0$, we consider the extreme points of the interval spanned by the
node's subtree. By doubling the universe size, we can assume these are
never elements of $S$ (alternatively, such extreme points are formal
rationals like $x + \frac{1}{2}$). We define $\overline{S}$ to be the
union of $S$ and the two special values for each branching node in the
primary trie; observe that $|\overline{S}| = O(n)$. We are interested
in holding $\overline{S}$ for navigation purposes: it gives a way to
find in constant time the maximum and minimum element from $S$ that
fits under a branching node (because these two values should be the
elements from $S$ closest to the special values for the branching
node).

\smallskip

Our data structure has the following components:

\begin{itemize}
\item[1.] a linked list with all elements of $S$ in increasing order,
  and a predecessor structure for $S$.
  
\item[2.] a linked list with all elements of $\overline{S}$ in
  increasing order, accompanied by a navigation structure which
  enables us to find in constant time the largest value from $S$
  smaller than a given value from $\overline{S} \setminus S$.  We also
  hold a predecessor structure for $\overline{S}$.
  
\item[3.] every branching node from the primary trie holds pointers to
  its lowest branching ancestor, and the two branching descendants
  (the highest branching nodes from the left and right subtrees; we
  consider leaves associated with elements from $S$ as branching
  descendants). We also hold pointers to the two extreme values
  associated with the node in the list in item 2. Finally, we hold a
  hash table with these branching nodes.

\item[4.] for each $t$, and every node $v$ in $T_t$, which is either a
  branching node or a child of a branching node on an active path, we
  hold the depth of the lowest branching ancestor of $r_0(v)$, using a
  Bloomier filter.
\end{itemize}

We begin by showing that this data structure takes linear space. Items
1-3 handle $O(n)$ elements, and have constant overhead per element.
We show below that the navigation structure from 2.~can be implemented
in linear space. The predecessor structure should also use linear
space; for van Emde Boas, this can be achieved through hashing
\cite{willard83predecessor}.

In item 4., there are $O(n)$ branching nodes per trie. In addition,
there are $O(n)$ children of branching nodes which are on active
paths. Thus, we consider $O(n\lg w)$ nodes in total, and hold $O(\lg
w)$ bits of information for each (a depth). Using our solution for the
Bloomier filter, this takes $O(n(\lg w)^2 + w)$ bits, which is $o(n)$
words. Note that storing the depth of the branching ancestor is just a
trick to reduce space. Once we have a node in $T_0$ and we know the
depth of its branching ancestor, we can calculate the ancestor in
$O(1)$ time (just ignore the bits below the depth of the ancestor). So
in essence these are ``compressed pointers'' to the ancestors.

We now sketch the navigation structure from item 2. Observe that the
longest run in the list of elements from $\overline{S} \setminus S$
can have length at most $2w$. Indeed, the leftmost and rightmost
extreme values for the branching nodes form a parenthesis structure;
the maximum depth is $w$, corresponding to the maximum depth in the
trie. Between an open and a closed parenthesis, there must be at least
one element from $S$, so the longest uninterrupted sequence of
parenthesis can be $w$ closed parenthesis and $w$ open parenthesis.

The implementation of the navigation structure uses classic ideas. We
bucket $\Theta(\sqrt{w})$ consecutive elements from the list, and then
we bucket $\Theta(\sqrt{w})$ buckets. Each bucket holds a summary
word, with a bit for each element indicating whether it is in $S$ or
not; second-order buckets hold bits saying whether first order buckets
contain at least one element from $S$ or not. There is also an array
with pointers to the elements or first order buckets. By shifting, we
can always insert another summary bit in constant time when something
is added. However, we cannot insert something in the array in constant
time; to fix that, we insert elements in the array on the next
available position, and hold the correct permutation packed in a word
(using $O(\sqrt{w} \lg w)$ bits). To find an element from $S$, we need
to walk $O(1)$ buckets. The time is $O(1)$ per traversed bucket, since
we can use the classic constant-time subroutine for finding the most
significant bit \cite{fredman93fusion}.

We also describe a useful subroutine, $\func{test-branching}(v)$,
which tests whether a node $v$ from some $T_t$ is a branching node. To
do that, we query the structure in item 4.~to find the lowest
branching ancestor of $r_0(v)$. This value is defined if $v$ is a
branching node, but the Bloomier filter may return an arbitrary result
otherwise. We look up the purported ancestor in the structure of item
3. If the node is not a branching node, the value in the Bloom filter
for $v$ was bogus, so $v$ is not a branching node. Otherwise, we
inspect the two branching descendants of this node. If $v$ is a
branching node, one of these two descendants must be mapped to $v$ in
the trie of order $t$, which can be tested easily.

\subsection{Implementation of Updates}

We only discuss insertions; deletions follow parallel steps
uneventfully. We first insert the new element in $S$ and
$\overline{S}$ using the predecessor structures. Inserting a new
element creates exactly one branching node $v$ in the primary trie.
This node can be determined by examining the predecessor and successor
in $S$. Indeed, the lowest common ancestor in the primary trie can be
determined by taking an xor of the two values, finding the most
significant bit, and them masking everything below that bit from the
original values \cite{alstrup01range}.

We calculate the extreme values for the new branching node $v$, and
insert them in $\overline{S}$ using the predecessor structure. Finding
the branching ancestor of $v$ is equivalent to finding the enclosing
parentheses for the pair of parentheses which was just inserted. But
$\overline{S}$ has a special structure: a pair of parentheses always
encloses two subexpressions, which are either values from $S$, or a
parenthesized expression (i.e., the branching nodes from $T_0$ form a
binary tree structure). So one of the enclosing parentheses is either
immediately to the left, or immediately to the right of the new
pair. We can traverse a link from there to find the branching
ancestor. Once we have this ancestor, it is easy to update the local
structure of the branching nodes from item 3. Until now, the time is
dominated by the predecessor structure.

It remains to update the structure in item 4. For each $t > 0$, we can
either create a new branching node in $T_t$, or the branching node
existed already (this is possible for $t > 0$ because nodes have many
children). We first test whether the branching node existed or not
(using the $\func{test-branching}$ subroutine). If we need to
introduce a branching node, we simply add a new new entry in the
Bloomier filter with the depth of the branching ancestor of $v$. It
remains to consider active children of branching nodes, for which we
must store the depth of $v$. If we have just introduced a branching
node, it has exactly two active children (if there exist more than two
children on active paths, the node was a branching node before). These
children are determined by looking at the branching descendants of
$v$; these give the two active paths going into $v$. Both descendants
are mapped to active children of the new branching node from $T_t$. If
the branching node already existed, we must add one active child,
which is simply the child that the path to the newly inserted value
follows. Thus, to update item 4., we spend constant time per $T_t$. In
total, the running time of an update is $T_{pred} + O(\lg w) = O(\lg
w)$.

\subsection{Implementation of Queries}

Remember that a query receives an interval $[a,b]$ and must return a
value in $S \cap [a,b]$, if one exists. We begin by finding the node
$v$ which is the lowest common ancestor of $a$ and $b$ in the primary
trie; this takes constant time \cite{alstrup01range}. Note that $v$
spans an interval which includes $[a,b]$. The easiest case is when $v$
is a branching node; this can be recognized by a lookup in the hash
table from item 3. If so, we find the two branching descendants of
$v$; call the left one $v_L$ and the right one $v_R$. Then, if $S \cap
[a,b] \ne \emptyset$, either the rightmost value from $S$ that fits
under $v_L$ or the leftmost value from $S$ that in fits under $v_R$
must be in the interval $[a,b]$. This is so because $[a,b]$ straddles
the middle point of the interval spanned by $v$. The two values
mentioned above are the two values from $S$ closest (on both sides) to
this middle point, so if $[a,b]$ is non-empty, it must contain one of
these two. To find these two values, we follow a pointer from $v_L$ to
its left extreme point in $\overline{S}$. Then, we use the navigation
structure from item 2., and find the predecessor from $S$ of this
value in constant time. The rightmost value under $v_R$ is the next
element from $S$. Altogether, the case when $v$ is a branching node
takes constant time.

Now we must handle the case when $v$ is not a branching node. If $S
\cap [a,b] \ne \emptyset$, it must be the case that $v$ is on an
active path. Below we describe how to find the lowest branching
ancestor of $v$, \emph{assuming that $v$ is on an active path}. If
this assumption is violated, the value returned can be arbitrary. Once
we have the branching ancestor of $v$, we find the branching
descendant $w$ which is in $v$'s subtree. Now it is easy to see, by
the same reasoning as above, that if $[a,b] \cap S \ne \emptyset$
either the leftmost or the rightmost value from $S$ which is under $w$
must be in $[a,b]$. These two values are found in constant time using
the navigation structure from item 2., as described above. So if
$[a,b] \cap S \ne \emptyset$, we can find an element inside $[a,b]$.
If none of these two elements were in $[a,b]$ it must be the case that
$[a,b]$ was empty, because the algorithm works correctly when $[a,b]
\cap S \ne \emptyset$.

It remains to show how to find $v$'s branching ancestor, assuming $v$
is on an active path, but is not a branching node. If for some $t >
0$, $v$ is mapped to a branching node in $T_t$, it will also be mapped
to a branching node in tries of higher order. We are interested in the
smallest $t$ for which this happens. We find this $t$ by binary
search, taking time $O(\lg\lg w)$. For some proposed $t$, we check
whether the node to which $v$ is mapped in $T_t$ is a branching node
(using the $\func{test-branching}$ subroutine). If it is, we continue
searching below; otherwise, we continue above.

Suppose we found the smallest $t$ for which $v$ is mapped to a
branching node. In $T_{t-1}$, $v$ is mapped to some $z$ which is
\emph{not} a branching node. Finding the lowest branching ancestor of
$v$ is identical to finding the lowest branching ancestor of $r_0(z)$
in the primary trie (since $z$ is a not a branching node, there is no
branching node in the primary trie in the subtree corresponding to
$z$). Since in $T_t$ $z$ gets mapped to a branching node, its natural
subtree in $T_{t-1}$ must contain at least one branching node. We have
two cases: either $z$ is the root or a leaf of the natural subtree
(remember that a natural subtree has two levels). These can be
distinguished based on the parity of $z$'s depth. If $z$ is a leaf,
the root must be a branching node (because there is at least another
active leaf). But then $z$ is an active child of a branching node, so
item 4.~tells us the branching ancestor of $r_0(z)$. Now consider the
case when $z$ is the root of the natural subtree. Then $z$ is above
any branching node in its natural subtree, so to find the branching
ancestor of $r_0(z)$ we can find the branching ancestor of the node
from $T_t$ to which the natural subtree is mapped. But this is a
branching node, so the structure in item 4.~gives the desired
branching ancestor. To summarize, the only super-constant cost is the
binary search for $t$, which takes $O(\lg\lg w)$ time.

\section{Tradeoffs from Dynamic Range Reporting}

Fix a value $B \in [2,\sqrt{w}]$; varying $B$ will give our tradeoff
curve.  For an arbitrary $t \in [0, \lg_B w]$, we define the trie of
order $t$ to be the trie of depth $w / B^t$ and alphabet of $B^t$
bits, which represents all numbers in $S$. We call the trie for $t =
0$ the primary trie. A node $v$ in a trie of order $t$ is represented
by a subtree of depth $B^t$ in the primary trie; we say that the root
of this subtree ``corresponds to'' the node $v$. A node from a trie of
order $t$ is represented by a subtree of depth $B$ in the trie of
order $t-1$; we call such a subtree a ``natural depth-$B$
subtree''. Alternatively, a depth-$B$ subtree is natural if it starts
at a depth divisible by $B$.

The root-to-leaf paths from the primary trie are boken into chunks of
length $B^t$ in the trie of order $t$. A trie of order $t$ is similar
to the $t$-th level (counted bottom-up) of the tree used for the
greater-than problem, since a path in the primary trie is seen as the
leaves of that tree. Indeed, every node on the $t$-th level of that
tree held information about a subtree with $B^t$ leaves; here one edge
in a trie of order $t$ summarizes a segment of length $B^t$ bits.
Also, a natural depth-$B$ subtree corresponds to $B$ siblings in the
old structure. On the next level, the $B$ siblings are contracted into
a node; in the trie of higher order, a natural depth-$B$ subtree is
also contracted into a node. 

Our data structure has the following new components:

\begin{itemize}
  
\item[5A.] choose this for the first branch of the tradeoff (faster
  updates, slower queries): hold the same information as in 4.~for
  each $t$, and every node $v$ in the trie of order $t$ which is not a
  branching node, is on an active path, and is the child of a
  branching node in the trie of order $t$.
  
\item[5B.] choose this for the second branch of the tradeoff: hold the
  same information as above for each $t$, and every node $v$ which is
  not a branching node, is on an active path, and has a branching
  ancestor in the same natural depth-$B$ subtree.
\end{itemize}

In item 5A., notice that for every $t$ there are at most $2n - 2$
children of branching nodes which are on active paths. We store $O(\lg
w)$ bits for each, and there are $O(\lg_B w)$ values of $t$, so we can
store this in a Bloomier filter with $o(n)$ words of space. In item
5B., the number of interesting nodes blows up by at most $B$ compared
to 5A., and since $B \leq \sqrt{w}$, we are still using $o(n)$ words
of space.

\paragraph{Updates.}
For each $t > 0$, we can either create a new branching node in the
trie of order $t$, or the branching node existed already. We first
test whether the branching node existed or not. If we just introduced
a branching node, it has at most two children which are not branching
nodes and are on active paths (if more than two such children exist,
the node was a branching node before). If the branching node was old,
we might have added one such child. These children are determined by
looking at the branching descendents of $v$ (these give the two active
paths going into $v$, one or both of which are new active paths going
into the node in the subtrie of order $t$). For such children, we add
the depth of $v$ in the structure from item 5A. If we are in case 5B,
we follow both paths either until we find a branching node, or the
border of the natural depth-$B$ subtree. For of these $O(B)$
positions, we add the depth of $v$ in item $5B$. To summarize, the
running time is $O(T_{pred} + \lg_B w)$ if we need to update 5A., and
$O(T_{pred} + B \lg_B w)$ is we need to update 5B.

\paragraph{Queries.}
We need to show how to find $v$'s branching ancestor, assuming $v$ is
on an active path, but is not a branching node. For some $t > 0$, and
all $t$'s above that value, $v$ will be mapped in the trie of order
$t$ to some branching node. That is the smallest $t$ such that the
depth-$B^t$ natural subtree containing $v$ contains some branching
node. We find this $t$ by binary search, taking time $O(\lg(\lg_B
w))$. For some proposed $t$, we check if the node to which $v$ is
mapped is a branching node in the trie of order $t$ (using the
subroutine described above). If it is, we continue searching below;
otherwise, we continue above.

Say we found the smallest $t$ for which $v$ is mapped to a branching
node. In the trie of order $t-1$, $v$ is mapped to some $w$ which is
not a branching node. Finding the lowest branching ancestor of $v$ is
identical to finding the lowest branching ancestor of the node
corresponding to $w$ in the primary trie (since $w$ is a not a
branching node, there is no branching node in the primary trie in the
subtree represented by $w$). In the trie of order $t$, $w$ gets mapped
to a branching node, so the natural depth-$B$ subtree of $w$ contains
at least one branching node. The either: (1) there is some branching
node above $w$ in its natural depth-$B$ subtree, or (2) $w$ is on the
active path going to the root of this natural subtree (it is above any
branching node).

We first deal with case (2). If $w$ is above any branching node in its
natural subtree, to find $w$'s branching ancestor we can find the
branching ancestor of the node from the trie of order $t$, to which
this subtree is mapped. But this is a branching node, so the structure
in item 4.~gives the branching ancestor $z$. We can test that we are
indeed in case (2), and not case (1), by looking at the two branching
descendents of $z$, and checking that one of them is strictly under
$v$.

Now we deal with case (1). If we have the structure 5B., this is
trivial. Because $w$ is on an active path and has a branching ancestor
in its natural depth-$B$ subtree, it records the depth of the
branching ancestor of the node corresponding to $w$ in the primary
trie. So in this case, the only super-constant cost is the binary
search for $t$, which is $O(\lg(\lg_B w))$. If we only have the
structure 5A., we need to walk up the trie of order $t-1$ starting
from $w$. When we reach the child of the branching node above $w$, the
branching node from the primary trie is recorded in item 5A. Since the
branching node is in the same natural depth-$B$ subtree as $w$, we
reach this point after $O(B)$ steps. One last detail is that we do not
actually know when we have reached the child of a branching node
(because the Bloomier filter from item 5A.~can return arbitrary
results for nodes not satisfying this property). To cope with this, at
each level we hope that we have reached the destination, we query the
structure in item 5A., we find the purported branching ancestor, and
check that it really is the lowest branching acestor of $v$. This
takes constant time; if the result is wrong, we continue walking up
the trie. Overall, with the structure of 5A.~we need query time
$O(\lg(\lg_B w) + B)$.

We have shown how to achieve the same running times (as functions of
$B$) as in the case of the greater-than function. The same calculation
establishes our tradeoff curve.

\section{Lower Bounds for the Greater-Than Problem}

A lower bound for the first branch of the tradeoff can be obtained
based on Fredman's proof idea \cite{fredman82sums}. We ommit the
details for now. To get a lower bound for the second case ($T_q <
O(\lg\lg n)$), we use the sunflower lemma of Erd\H{o}s and Rado. A
sunflower is collection of sets (called petals) such that the
intersection of any two of the sets is equal to the intersection of
all the sets.

\begin{lemma}[Sunflower Lemma]
  Consider a collection of $n$ sets, of cardinalities at most $s$. If
  $n > (p-1)^{s+1} s!$, the collection contains as a subcollection a
  sunflower with $p$ petals.
\end{lemma}

For every query parameter in $[0,n-1]$, the algorithm performs at most
$T_q$ probes to the memory. Thus, there are $2^{T_q}$ possible
execution paths, and at most $2^{T_q} - 1$ bit cells are probed on at
least some execution path. This gives $n$ sets of cells of sizes at
most $s = O(2^{T_q})$; we call these sets query schemes. By the
sunflower lemma, we can find a sunflower with $p$ petals, if $p$
satisfies: $n > (p-1)^{s+1} s! \Rightarrow \lg n > \Theta(s (\lg p +
\lg s))$. If $T_q < (1-\epsilon) \lg\lg n$, we have $s\lg s = o(\lg
n)$, so our condition becomes $\lg n > \Theta(s \lg p)$. So we can
find a sunflower with $p$ petals such that $\lg p = \Omega((\lg
n)/s)$. Let $P$ be the set of query parameters whose query schemes are
these $p$ petals.

The center of the sunflower (the intersection of all sets) obviously
has size at most $s$. Now consider the update schemes for the numbers
in $P$. We can always find $T \subset P$ such that $|T| \geq |P| /
2^s$ and the update schemes for all numbers in $T$ look identical if
we only inspect the center of the sunflower. Thus $\lg |T| = \lg |P| -
s = \Omega(\frac{\lg n}{s} - s)$. If $T_u < (\frac{1}{2} - \epsilon)
\lg\lg n$, we have $s = o(\frac{\lg n}{s})$, so we obtain $\lg |T| =
\Omega(\frac{\lg n}{s})$.

Now we restrict our attention to numbers in $T$ for both the update
and query value. The cells in the center of the sunflower are thus
fixed. Define the natural result of a certain query to be the result
(greater than vs. not greater than) of the query if all bit cells read
by the query outside the center of the sunflower are zero. Now pick a
random $x \in T$. For some $y$ in the middle third of $T$ (when
considering the elements of $T$ in increasing order), we have $\Pr[y
\leq x] \geq \frac{1}{3}, \Pr[y > x] \geq \frac{1}{3}$, so no matter
what the natural result of querying $y$ is, it is wrong with
probability at least $\frac{1}{3}$. So for a random $x$, at least a
fraction of $\frac{1}{9}$ of the natural results are wrong. Consider
an explicit $x$ with this property. The update scheme for $x$ must set
sufficiently many cells to change these natural results. But these
cells can only be in the petals of the queries whose natural results
are wrong, and the petals are disjoint except for the center, which is
fixed. So the update scheme must set at least one cell for every
natural result which is wrong. Hence $T_u \geq |T|/9 \Rightarrow \lg
T_u = \Omega(\lg |T|) = \Omega(\frac{\lg n}{s}) = \Omega(\frac{\lg
  n}{2^{T_q}}) \Rightarrow 2^{T_q} = \Omega(\lg_{T_u} n)$.

\paragraph*{Acknowledgement.}  
The authors would like to thank Gerth Brodal for discussions in the
early stages of this work, in particular on how the results could be
extended to dynamic range counting.

\bibliographystyle{plain}
\bibliography{../general}


\end{document}